\documentclass[prl,showpacs,amsmath,twocolumn]{revtex4}  
\newcommand{\beq}{\begin{equation}}
\newcommand{\enq}{\end{equation}}

\usepackage{graphicx}

\begin{document}
\title{Collective stimulated Brillouin backscatter}

\author{Pavel M. Lushnikov$^{1}$ and Harvey A. Rose$^2$
}

\affiliation{$^1$ Department of Mathematics and Statistics, University of New Mexico, Albuquerque, NM 87131, USA \\
$^2$Theoretical Division, Los Alamos National Laboratory,
  MS-B213, Los Alamos, New Mexico, 87545
}

\email{har@lanl.gov}

\date{
\today
}

\begin{abstract}
We develop the statistical theory of the stimulated Brillouin backscatter (BSBS) instability of a spatially and temporally partially incoherent laser beam for laser fusion relevant plasma. We find
a new regime of BSBS which has a much larger threshold than the classical threshold of a coherent beam in long-scale-length laser fusion plasma. Instability is collective because it does not depend on
the dynamics of isolated speckles of laser intensity, but rather depends on averaged beam intensity. We identify convective and absolute instability regimes.  Well above the incoherent threshold the coherent instability growth rate is recovered. The  threshold of
convective instability is inside the typical parameter region of National Ignition Facility (NIF) designs although current NIF bandwidth is not large enough to insure dominance of collective instability and suggests lower instability threshold due to speckle contribution.
In contrast, we estimate that the bandwidth of  KrF-laser-based fusion systems would be large enough.
\end{abstract}


\maketitle

Inertial confinement fusion (ICF) experiments require propagation of intense laser light through underdense plasma subject to laser-plasma instabilities which can be deleterious for
achievement of thermonuclear target ignition because they can cause the loss of target symmetry and hot electron production \cite{Lindl2004}.
Among laser-plasma instabilities, the backward stimulated Brillouin backscatter (BSBS) has been considered  for a long time as serious danger because the damping threshold of BSBS of coherent laser beams is  typically several order of magnitude lower compared with the required laser intensity $\sim  10^{15}\mbox{W}/\mbox{cm}^2$ for ICF. Recent experiments for a first time achieved conditions of fusion plasma and indeed demonstrated that large levels of BSBS (up to tens percent of reflectivity) are possible\cite{FroulaPRL2007}.

Theory of laser-plasma interaction (LPI) instabilities is well developed for coherent laser beam \cite{Kruer1990}.  However, ICF laser beams are not coherent because temporal and spatial beam smoothing techniques are currently used to produce laser beams with short correlation time, $T_c,$  and lengths to suppress laser-plasma interactions. The laser intensity forms a speckle field - a random in space distribution of intensity with transverse correlation length
$l_c\simeq 2F/k_0$ and longitudinal correlation length (speckle length) $L_{speckle}\simeq 7F^2\lambda_0$, where $F$ is the optic $f/\#$ and $\lambda_0=2\pi/k_0$ is the wavelength  (see e.g. \cite{RosePhysPlasm1995,GarnierPhysPlasm1999}). Beam smoothing is a part of most constructed and suggested ICF facilities. However, instability theory of  smoothed laser beam interaction with plasma is not well developed. There are intense experimental and simulation ongoing efforts \cite{FroulaPRL2007} to determine BSBS threshold for smoothed beams which appears to be in some cases quite low so that it is now under discussion that laser intensity at the National Ignition Facility (NIF) should lowered by a factor of few compared with  original NIF designs\cite{Lindl2004} with intensities  $\sim 2\times 10^{15}\mbox{W}/\mbox{cm}^2$.
\begin{figure}
\begin{center}
\includegraphics[width = 2.3 in]{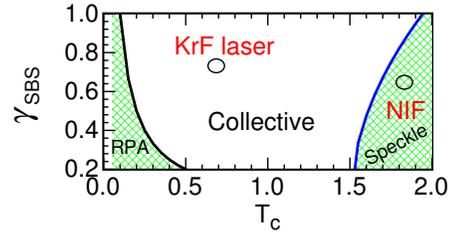}
\end{center}
\caption{ Regimes of BSBS of partially incoherent laser beam. $T_c$ is the laser correlation time and $\gamma_{BSBS}=\omega_i$ is the BSBS temporal grow rate, both in arbitrary units. For large $T_c$ instability is dominated by speckles of laser field (on the right). CBSBS regime is shown in the middle. Very small $T_c$ corresponds to RPA regime (on the left). Solid lines sketch transitions between all 3 regimes.
}
\label{fig:fig0}
\end{figure}

Here we develop a theory of collective BSBS instability (CBSBS), which is a new BSBS regime, for propagation of laser beam with finite $T_c$  in homogeneous plasma. CBSBS has threshold comparable with  NIF intensities.
CBSBS requires  $T_c$ small enough to suppress contribution from speckles. If we additionally assume that $T_c\gg L_{speckle}/c$ then CBSBS  threshold does not depend on $T_c$. Such $T_c$ is accessible to  KrF lasers \cite{Weaver2007}, $T_c\simeq 0.7\mbox{ps}$, but not for NIF glass lasers with beam smoothing up to 3{\AA} at $1\omega$, implying $T_c\simeq 4\mbox{ps}$ at  $3\omega$.  This is consistent with the numerical simulations which show that  BSBS threshold in NIF emulation experiments is dominated by speckles \cite{FroulaPRL2007,BergerPrivate2007}. We show below that speckle-dominated threshold is lower by a factor 7 than CBSBS threshold. Since plasma inhomogeneity can only increase  instability threshold \cite{Kruer1990}, The CBSBS threshold  is a lower bound.  Fig. \ref{fig:fig0} depicts
CBSBS between large $T_c$ speckle regime \cite{RoseDuBois1994} and random phase approximation (RPA) \cite{vedenov1964,DuBoisBezzeridesRose1992,PesmeBerger1994} regime.

Assume that laser beam propagates in plasma with frequency
$\omega_0$  along $z$ with the electric field
$\cal E$ given by
\begin{eqnarray}\label{EBdef}
{\cal E}=(1/2)e^{-i\omega_0 t}\Big [E e^{ik_0 z}+Be^{-ik_0
z-i\Delta\omega t}\Big ]+c.c.,
\end{eqnarray}
where $E({\bf r}, z,t)$ is the envelope of laser beam and $B({\bf
r}, z,t)$ is the envelope of backscattered wave,  ${\bf r}=(x,y)$,
and c.c. means complex conjugated terms. Frequency shift $\Delta
\omega=-2k_0c_s$ is determined by coupling of $E$ and $B$ through ion-acoustic wave with phase speed
$c_s$ and wavevector $2k_0$ with  plasma density fluctuation
$\delta n_e$ given by $\frac{\delta n_e}{n_e}=\frac{1}{2}\sigma
e^{2ik_0z+i\Delta\omega t}+c.c.,$ where $\sigma({\bf r}, z,t)$ is
the slow envelope and $n_e$ is the average electron density, assumed
to be small compared to critical density, $n_c$. The coupling of $E$
and $B$ to plasma density fluctuations gives, ignoring light wave
damping,
\begin{eqnarray}\label{EBeq1}
  \left [ i\Big (c^{-1}{\partial_t}+{\partial_z}\Big )+(2k_0)^{-1}\nabla^2
  \right ]E=\frac{k_0}{4}\frac{n_e}{n_c}\sigma B, \\
  \left [ i\Big (c^{-1}{\partial_t}-{\partial_z}\Big )+(2k_0)^{-1}\nabla^2
  \right ]B=\frac{k_0}{4}\frac{n_e}{n_c}\sigma^* E, \label{EBeq2}
\end{eqnarray}
 $\nabla=({\partial_x},{ \partial_y})$, and $\sigma$ is described by the acoustic wave equation coupled to
the pondermotive force $\propto {\cal E}^2$ which results in the
envelope equation
\begin{eqnarray}\label{sigma1}
   [ i ({c_s^{-1}}{\partial_t}+2\nu_{ia} k_0+{\partial_z} )-(4k_0)^{-1}\nabla^2
   ]\sigma^*=-2k_0  E^*B,
\end{eqnarray}
where we neglected terms $\propto |E|^2, \ |B|^2$ in r.h.s. which are responsible for self-focusing effects, $\nu_L$ is the Landau damping of ion-acoustic wave and $\nu_{ia}=\nu_L/2k_0c_s$ is the scaled acoustic damping coefficient.  $E$ and $B$ are in thermal units (see e.g. \cite{LushnikovRosePRL2004}).

Assume that laser beam was made partially incoherent through induced spacial incoherence beam smoothing \cite{LehmbergObenschain1983} which defines stochastic boundary conditions at $z=0$ for the
spacial Fourier transform (over ${\bf r}$) components $ \hat E({\bf k})$, of laser beam amplitude
\cite{LushnikovRosePRL2004}:
\begin{eqnarray}\label{phik}
\hat E({\bf k },z=0,t)= |E_{\bf k}|\exp [ i\phi_{\bf
k}(t) ], \nonumber \\
  \langle \exp i [\phi_{\bf
k}(t)-\phi_{{\bf k}'}(t') ] \rangle =\delta_{{\bf k
k}'}\exp (-|t-t'|/T_c),
 \nonumber \\
 |E_{\bf k}|=const, \ k<k_m; \  E_{\bf k}=0, \ k>k_m,
\end{eqnarray}
chosen to  the idealized "top hat" model of NIF optics \cite{polarization}. Here $k_m\simeq k_0/(2F)$ and the average intensity, $   \langle I
\rangle \equiv  \langle |E|^2 \rangle =I$ determines the constant.

In linear approximation, assuming $|B|\ll |E|$ so that only laser beam is BSBS unstable, we can neglect  right hand side (r.h.s.) of Eq. (\ref{EBeq1}). The
resulting linear equation with top hat boundary condition (\ref{phik}) has the exact solution as
 decomposition of $E$ into
 Fourier series,
$E({\bf r},z,t)=\sum_j E_{{\bf k}_j}$
%
with $E_{{\bf k}_j} \propto \exp\big [ i(\phi_{{\bf k}_j}(t-z/c)+{\bf k}_j\cdot {\bf r}-{\bf k}_j^2z/2k_0)\big ].$ Eq. (\ref{EBeq2}) is linear in $B$ and $E$  which implies that
$B$ can be also decomposed into $B=\sum_j B_{{\bf k}_j}.$
We approximate r.h.s. of (\ref{sigma1}) as $E^*B\simeq \sum_j
E_{{\bf k}_j}^*
B_{{\bf k}_j}$ so that %
\begin{eqnarray}\label{sigma2}
  \left [ i (c_s^{-1}{\partial_t}+2\nu_{ia} k_0+{\partial_z})-(4k_0)^{-1}\nabla^2
  \right ]\sigma^*\nonumber \\ =-2k_0  \sum_jE_{{\bf k}_j}^*B_{{\bf k}_j},
\end{eqnarray}
which means that we neglect off-diagonal terms $E_{{\bf k}_j}^*
B_{{{\bf k}_j}'}, \quad j\neq j'.$ Since speckles of
laser field arise from interference of different Fourier modes,
$j\neq j',$ we associate the off-diagonal terms with speckle
contribution to BSBS (independent hot spot model
\cite{RoseDuBois1993,RosePhysPlasm1995}). Speckle contribution can
can be neglected if \cite{MounaixPRL2000}
\begin{eqnarray}\label{TcTsaturation}
T_c \ll t_{sat},
\end{eqnarray}
where $t_{sat}$ is the characteristic time scale at which BSBS convective gain saturates.

We use the linear part of the theory of Ref. \cite{MounaixPRL2000} to estimate $T_{sat}$ for speckle contribution to backscatter as $t_{sat}=(L_{speckle}/c)\big[2+(\gamma_0/\nu_L)^2\big ]$,
where
 $\gamma_0^2=k_0^2 c c_s I_{speckle}n_e/2n_c$ and we choose the
typical intensity of light in speckle $I_{speckle}=3I$, where $I$ is the spatial average of laser intensity $|E|^2$ \cite{RoseDuBois1993,GarnierPhysPlasm1999,MounaixPRL2000}. In such a case $T_c
/t_{sat}\simeq T_ck_0c_s\nu_{ia}/(4 \tilde I)$, where here and below $\tilde I$ designates the scaled dimensionless laser intensity defined as   $\tilde I=\frac{4F^2}{\nu_{ia}}\frac{n_e}{n_c} I$. For
typical NIF parameters $\tilde I\sim 1$ \cite{Lindl2004,LushnikovRosePlasmPhysContrFusion2006}, $\lambda_0=351 \mbox{nm}$ and $c_s=6\times 10^{7}\ \mbox{cm s}^{-1}$   we obtain from (\ref{TcTsaturation}) the estimate $T_c \ll 0.4/\nu_{ia}\ \mbox{[ps]}$
which is not satisfied for low plasma ionization number $Z$ plasma in NIF which typically has $\nu_{ia}\sim 0.1$.  However, CBSBS can still be relevant for NIF in gold plasma near hohlraum \cite{Lindl2004} with $\nu_{ia}\sim 0.01$. Similar estimate for KrF lasers ($\lambda_0=248 \mbox{nm}, \ F=8$) gives $T_c \ll 0.3/\nu_{ia}\ \mbox{[ps]}$  which is easier to satisfy because of smaller $T_c$ and suggests that KrF lasers  are better suited for applicability of CBSBS.

If we look for solution of Eqs. (\ref{EBeq2}) and (\ref{sigma2}) in exponential form $B_j, \sigma^* \propto e^{i(\kappa z+{\bf k}\cdot {\bf r}-\omega t)}$, we arrive at the following dispersion
relation in dimensionless units
\begin{eqnarray}\label{dispk1}
-i\omega+\mu+i\kappa-(i/4)k^2\nonumber \\
=8iF^4\frac{n_e}{n_c}\sum\limits_{j=1}^N \frac{|E_j|^2}{\omega\frac{c_s}{c}+\kappa-k_j^2-\frac{k^2}{2}-{\bf k}_j\cdot {\bf k}},
\end{eqnarray}
where $\mu\equiv 2\nu_{ia} k_0^2/k_m^2,$ $1/k_m$ is the transverse unit of length, $k_0/k_m^2$ is the unit in $z$ direction, $k_0/k_m^2 c_s$ is the time unit and $I=\sum_j|E_{{\bf k}_j}|^2$.

The dispersion relation (\ref{dispk1}) is correct provided the temporal growth rate $\omega_i=Im(\omega)$ is small compare to inverse time of light propagation along speckle,
$\omega_i\ll c/L_{spekle}$, and if during  time $T_c$ light travels much further than a speckle length, $L_{speckle}\ll cT_c$. That second condition ensures that term
$\propto\phi'_{{\bf k}_j}(t-z/c)\sim 1/T_c$ could be neglected in Eq. (\ref{EBeq2}) allowing the time dependence of $E$ in  Eqs. (\ref{EBeq2}) and (\ref{sigma2}) to be ignored and in such case density fluctuation $\sigma$ evolves without fluctuations.
 E.g. for typical  NIF parameters, $T_ck_0c_s/2F\sim 1$ we obtain that $2c/7c_sF\gg 1$ which is well
satisfied for NIF optics \cite{Lindl2004}.

In the continuous limit $N\to \infty$, sum in (\ref{dispk1}) is replaced by integral which gives for most unstable mode ${\bf k}=0$:
\begin{eqnarray}\label{dispk0cont}
\Delta(\omega,\kappa)=-i\omega+\mu+i\kappa+i\frac{\mu}{4}\tilde I\ln\frac{1-\kappa-\omega\frac{c_s}{c}}{-\kappa-\omega\frac{c_s}{c}}
=0.
\end{eqnarray}

Eq.  (\ref{dispk0cont}) has branch cut in complex $\kappa$ plane determined by points $\kappa_1=1-\omega\frac{c_s}{c}$ and $\kappa_2=-\omega\frac{c_s}{c}$. Standard analysis
of convective vs. absolute instabilities (see e.g. \cite{Briggs1964}) should be modified to include that branch cut. In discrete case with $N \gg 1$ instead of branch cut the discrete dispersion
relation (\ref{dispk1}) has solutions located near the line $(\kappa_1,\kappa_2)$. These solutions are highly localized around some ${\bf k}_j$ so they cannot be approximated by (\ref{dispk0cont})
but they are stable for  $N \gg 1$. Generally there are two solutions of (\ref{dispk0cont}), however for $Im(\omega)\to \infty$ one solution is absorbed into
branch cut. Second solution is stable. Above the convective CBSBS threshold,
\begin{eqnarray}\label{I0convthresh}
\tilde I_{convthresh}=4/\pi,
\end{eqnarray}
the first solution crosses real $\kappa$ axis from below as $Im(\omega)\to 0$ so it describes instability of backscattered wave with $Im(\kappa)>0.$

However, above the absolute CBSBS threshold, which can be approximated from solution of Eq. (\ref{dispk0cont}) as
\begin{eqnarray}\label{I0absthresh}
\tilde I_{absthresh}\simeq(1/2)\Big (\mu^{-1}+\mu+\sqrt{\mu^2-2}\Big ), \quad 
 \mu\gtrsim 4,
\end{eqnarray}
the contour $Im(\omega)=Const$ cannot be moved down to real $\omega$ axis because of pinching of two solutions of  (\ref{dispk0cont}) which defines growth rate of absolute
instability. We conclude that classical analysis of instabilities still holds for incoherent beam if we additionally allow the absorption of one solution branch into branch cut. This effect
results from incoherence of pump beam which has infinitely many transverse Fourier modes in approximation of Eq. (\ref{dispk0cont}) and there is no counterpart of that effect for coherent beam.

For $\mu\gg 1  $ the absolute threshold (\ref{I0absthresh}) reduces to the coherent absolute BSBS instability threshold
\begin{eqnarray}\label{I0absthreshcoherent}
\tilde I_{absthreshcoherent}= \mu.
\end{eqnarray}

\begin{figure}
\begin{center}
\includegraphics[width = 1.9 in]{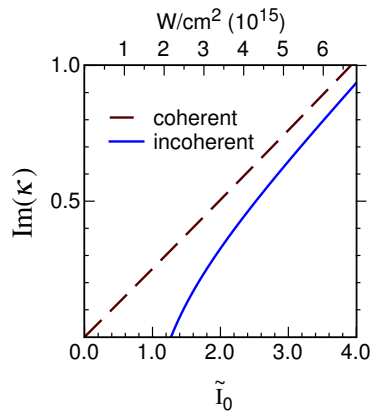}
\end{center}
\caption{ Convective instability gain rate $Im(\kappa)$ as a function of dimensionless laser beam intensity (lower grid) for incoherent laser beam (solid curve) and coherent beam (dashed curve) for
$\mu=51.2$. Here $\mu=8F^2\nu_{ia}$. Upper grid corresponds to laser intensity in dimensional units for typical NIF parameters  $T_e\simeq 5\mbox{keV}, \ F=8,\  \ n_e/n_c=0.1, \nu_{ia}=0.1, \ \omega_0\approx3.6\times 10^{15}\mbox{sec}^{-1}$.
Note that in the range between $\nu_{ia}=0.05$ and  $\nu_{ia}=0.3$ there is no qualitative change in the behavior of both curves except change of scale at upper axis. }
\label{fig:fig1}
\end{figure}

For  NIF parameters, $T_e\simeq 5\mbox{keV}, \ F=8,\  \ n_e/n_c=0.1, \ \lambda_0=351 \mbox{nm}$ with moderate acoustic damping, $\nu_{ia}\simeq 0.1$,
we obtain in dimensional units $I_{convthresh}\simeq 2\times 10^{15}\mbox{W}/\mbox{cm}^2$ and $I_{absthresh}\simeq 9\times 10^{16}\mbox{W}/\mbox{cm}^2.$ For high $Z$ plasma (e.g. gold
plasma near the wall of NIF hohlraum \cite{Lindl2004}, $\nu_{ia}\simeq 0.01$) we obtain $I_{convthresh}\simeq 2\times 10^{14}\mbox{W}/\mbox{cm}^2$ and $I_{absthresh}\simeq 9\times
10^{14}\mbox{W}/\mbox{cm}^2.$ Typical intensity of NIF laser shots is between $10^{15}\mbox{W}/\mbox{cm}^2$ and $2 \times 10^{15}\mbox{W}/\mbox{cm}^2$ so we conclude that in different parts of NIF
plasma both convective and absolute instabilities are possible. Fig. \ref{fig:fig1} compares instability gain rate of coherent and incoherent beams for $\mu=51.2$.

In contrast with Eq. (\ref{I0convthresh}), the convective instability threshold in coherent case is 0 because we neglect damping of $B$ in Eq. (\ref{EBeq2}). Retaining collisional light damping gives finite threshold  $\tilde I_{convcoherent}=16F^2 \nu_B/k_0 c\ll 1$, where $\nu_B=\frac{n_e}{n_c}\frac{\nu_{ei}}{2}$ \cite{Kruer1990} is the collisional damping of backscattered wave $B$ and $\nu_{ei}$ is the electron-ion collision frequency.  That threshold is  several orders of magnitude smaller compared with (\ref{I0convthresh}) and is neglected here.  Qualitatively incoherence of laser beam can be considered as effective damping
of $B$  with  effective damping rate $\nu_{effective}=\frac{\pi k_0 c_s}{16F^2}$.

\begin{figure}
\begin{center}
\includegraphics[width = 1.9 in]{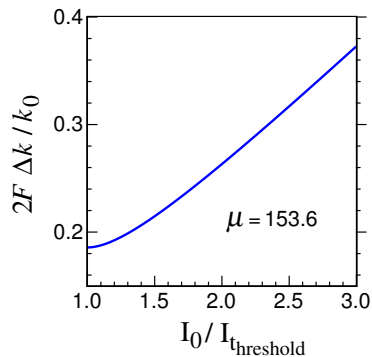}
\end{center}
\caption{Angular width $2F\triangle k/k_0$ of  convective instability gain rate $Im(\kappa)$ vs. laser intensity scaled to (\ref{I0convthresh})  for
$\mu=153.6$ There is no qualitative change in that curve for smaller $\mu$ down to $\mu=5.12$. For NIF optics,  $F=8,$ value $\mu=153.6$ corresponds to $\nu_{ia}=0.3$ (e.g. He-H plasma) and  $\mu=5.12$ corresponds to
 $\nu_{ia}=0.01$ (e.g. gold plasma).}
\label{fig:fig2}
\end{figure}

Depending on laser incoherence we have a hierarchy of thresholds:
(a)Spatially incoherent  laser beam with large $T_c$ has threshold, $\tilde I_{threshold}=\tilde I_{thresholdspeckle}=4/7\pi$ which is dominated by intense speckles \cite{RoseDuBois1994}.
(b)Spatial and temporary incoherent beam with $T_c$ satisfying (\ref{TcTsaturation}) is given by (\ref{I0convthresh}) which factor $7$ times higher compared with speckle threshold and does not depends on $T_c.$
It indicates practical limit of how threshold of BSBS instability can be increased by decreasing $T_c.$
(c)For much smaller $T_c$, such that it is smaller than both inverse acoustic damping $T_c\ll 1/\nu_L$ and inverse temporal growth rate $T_c\ll 1/\omega_i$, the classical RPA regime is recovered which has ignorable  diffraction  (\cite{vedenov1964,DuBoisBezzeridesRose1992,PesmeBerger1994}). This limit (e.g. for $\lambda_0=351\mbox{nm}$ and $\nu_{ia}=0.15$ it requires $T_c\ll 0.3\mbox{ps}$) is not practical for ICF as $T_c$ is too small.  Cases (a)-(c) are shown in Fig. \ref{fig:fig0}.

Current NIF 3{\AA} beam smoothing design  is between regimes (a) and (b). KrF laser  with $T_c\simeq 0.7\mbox{ps}$ would be in regime (b). Thus generally we expect that KrF-laser-based ICF allows access to CBSBS regime  although CBSBS threshold for NIF can be possibly initiated by self-induced temporal incoherence  (see e.g. \cite{SchmittAfeyan1998}).
Another possibility for self-induced temporal incoherence is through collective forward stimulated Brillouin scatter (CFSBS) instability \cite{LushnikovRosePRL2004,LushnikovRosePlasmPhysContrFusion2006}. Above CFSBS threshold correlation length decreases with beam propagation length and may decrease $T_c$. For low $Z$ plasma threshold for CFSBS is close to (\ref{I0convthresh}) \cite{LushnikovRosePRL2004}. As $Z$ increases (which can be achieved by adding high $Z$ dopant), CFSBS threshold decreases below (\ref{I0convthresh}) and  might result in decrease of $T_c.$

To distinguish contribution to BSBS from speckles (regime (a)) and CBSBS (regime (b)) we propose to look at angular divergence $\triangle \theta=\triangle k/k_0$ of BSBS.   In general one expects gain narrowing of the scattered light: the modes close to the most unstable mode, with gain rate $(\kappa_i)_{max}$,  dominate. Here $\kappa_i \equiv Im(\kappa)$. Fig. \ref{fig:fig2} shows $2F\triangle \theta$ from CBSBS as a function of laser intensity above CBSBS threshold at propagation distance $L=10/ (\kappa_i)_{max}$. $L$ is chosen from the physical condition that there is sufficient convective CBSBS gain, to amplify the energy of thermal acoustic fluctuations at wavenumber $2k_0$ to have
reflectivity $\sim 1$, and for fusion plasma this is typically  $\exp(G)=\exp(20)$ (see e.g. \cite{FroulaPRL2007}), where $G=2\kappa_i L$ is the power convective gain exponent.    Then $\triangle \theta$ is conventionally defined by half width at half maximum:   $\exp[G(\triangle \theta)] = 0.5  \exp[G(\theta)]|_{\theta=0}$. Important feature of CBSBS seen in Fig. \ref{fig:fig2} is that $\triangle\theta\neq 0$ at threshold with $\kappa_i( k)\simeq\kappa_i(0)(1-\tilde \alpha k^2)$ and $\tilde \alpha\simeq\mu\tilde I/(\mu\tilde I-1)$ near threshold. Fig. \ref{fig:fig2} should be compared with $\triangle \theta$ from speckle-dominated backscatter.  Previous work \cite{DivolMounaixPRE1998} suggested that speckles can also cause
$\triangle \Theta$ below top-hat width, $1/2F$, for very intense speckle backscatter.  We estimate based on Refs. \cite{RosePhysPlasm1995,GarnierPhysPlasm1999}) that for nominal ICF plasma ($\sim 10^5$ speckle volumes), most intense speckle  is $\sim 15 I$  which gives  $G_{intense}= 15 \langle G_0\rangle \simeq 100$ near CBSBS threshold, where $\langle G_0\rangle =2\kappa_i L_{speckle }\simeq 7$ is the the gain over speckle with the average intensity $I.$ We performed direct simulations of backscatter from $G_{intense}= 100$ speckle and found that $\triangle \theta\simeq 1/2F$ which means that asymptotic \cite{DivolMounaixPRE1998} is still not applicable. In other words, finite size plasma effects dominates over asymptotic theory of infinite plasma.  We conclude that regime (a) can be easily distinguished from CBSBS regime (b):  near CBSBS threshold with condition (\ref{TcTsaturation}) satisfied one should see backscattered light spectrum with essential peak whose width is given by Fig.   \ref{fig:fig2} and wide weak background determined by speckles.

In summary, we found a novel coherence time regime in which $T_c$ is too large for applicability of well-known statistical theories (RPA) but rather an intermediate regime, $T_c$ is
small enough to  suppress  speckle BSBS. Unlike coherent beam CBSBS has threshold typically much larger than
that determined by damping while for laser intensity many
times above convective instability threshold for incoherent beam,
the coherent theory is recovered.


We acknowledge helpful discussions with B. Afeyan, R. Berger, L. Divol, D. Froula, and N. Meezan.
This work was carried out under the auspices of the NNSA of the DOE at LANL under Contract No. DE-AC52-06NA25396




\end{document}